\documentclass[twocolumn]{openjournal}
\usepackage{graphicx}   
\usepackage{amsmath}    
\usepackage{amssymb,amsfonts}    
\usepackage{orcidlink}
\usepackage{verbatim}
\usepackage{hyperref}
\usepackage{xcolor}
\usepackage{longtable}
\usepackage{booktabs}
\usepackage{capt-of}

\shorttitle{A white dwarf companion to Nashira=Gamma Capricorni}
\shortauthors{Idel Waisberg \& Boaz Katz}

\begin{document}

\title{A lucky one indeed:\\An elusive white dwarf companion to Nashira=Gamma Capricorni}

\footnotetext[]{Based on observations collected at the European Southern Observatory, Chile, Program ID 114.27DS.001}

\newcommand{\weizmann}{Department of Particle Physics and Astrophysics, Weizmann Institute of Science, Rehovot 76100, Israel}

\email{email: idelwaisberg@gmail.com}

\author{\vspace{-1.2cm}Idel Waisberg\,\orcidlink{0000-0003-0304-743X}$^{1}$ \& Boaz Katz\orcidlink{0000-0003-0584-2920}$^{2}$}

\affiliation{$^1$Independent researcher, Lambda Ophiuchi Ltda}
\affiliation{$^2$\weizmann}

\begin{abstract}
Nashira=Gamma Capricorni ($\gamma$ Cap) is a bright Am star that was identified as a $P\sim800\text{ days}$ astrometric binary in the original Hipparcos reduction. Using radial velocities, we confirm that $\gamma$ Cap is a $P=840 \text{ days}$ single-lined spectroscopic binary. From a combined fit with the Hipparcos 1997 Intermediate Astrometric Data and using the evolutionary mass of the subgiant primary $M_A \simeq 2.06\pm0.08 M_{\odot}$, we find a companion mass $M_B = 0.56 \pm 0.06 M_{\odot}$ and a semi-major axis $a=2.40 \text{ au} \leftrightarrow 58 \text{ mas}$. A VLTI/GRAVITY interferometric observation did not detect the companion and safely excludes a red dwarf of this mass, implying that it is very likely a white dwarf. Post-AGB tidal circularization naturally accounts for the low eccentricity $e < 0.1$ and mass transfer from the former AGB companion helps to explain the extreme s-process enrichment of Nashira (which stands out even among other Am stars) as well as its solar carbon abundance, suggesting that it is a hybrid Am and barium star. Nashira is among the five visually brightest stars with a white dwarf companion together with Sirius, Procyon, Tianjinnan and Mira (at its visual peak). 
\end{abstract}


\section{Introduction}
\label{sec:introduction}

Two among ten of the visually brightest stars in the night sky (namely, Sirius and Procyon) have white dwarf (WD) companions. While such a high rate of WD companions to the very closest A- and F-type stars is a fluke, the multiplicity statistics of intermediate-mass stars do imply that there should be a significant number of undiscovered white dwarf companions among the local stellar population \citep[e.g.][]{Holberg2013,Waisberg2023}. WD companions to A and F-type stars are particularly difficult to discover due to the extreme contrast ratios involved. Except for the cases of very nearby stars or very wide orbits in which the WD can be spatially resolved, only in the ultraviolet bands there is a chance to directly detect an unresolved WD companion (and even then only for the minority of WDs that are hot enough). Unresolved and cooler WD companions to such stars cannot currently be directly detected. 

A companion with a mass consistent with a WD can still be inferred from radial velocity and/or photocenter astrometry, but it is difficult to exclude the possibility of a red dwarf companion of similar mass since it can also be quite faint compared the primary star (albeit orders of magnitude brighter than a WD) and thereby difficult or impossible to detect spectroscopically. Near-infrared (NIR) interferometry is particularly useful for differentiating between these two cases: while the detection of a WD companion is orders of magnitude beyond the current sensitivity of NIR interferometers, the typical NIR flux ratio $\sim 1\%$ for a $M\gtrsim 0.5 M_{\odot}$ red dwarf companion to an A-type star is safely within the detection limit of current instruments such as VLTI/GRAVITY \citep{GRAVITY17}.

\begin{figure*}[]
\includegraphics[width=2\columnwidth]{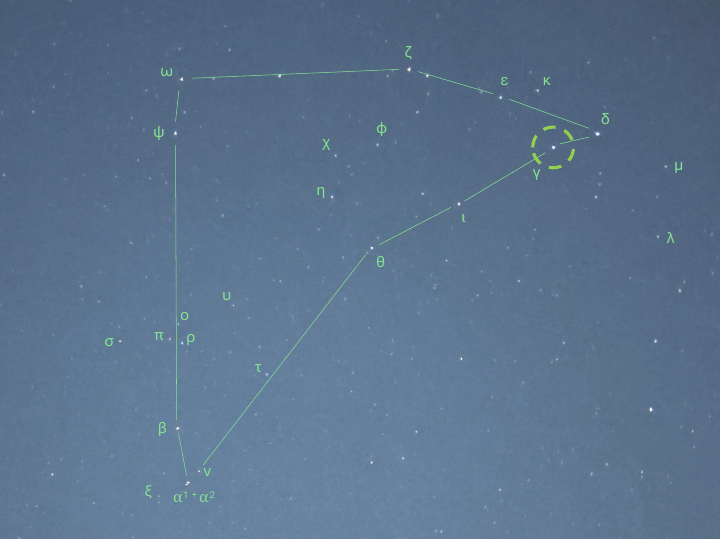}
\caption{\label{fig:sky_image}  Smartphone photo of Capricornus as seen from São Roque, SP, Brazil on 2026-07-10 UTC 04:43. In this paper we show that Nashira=Gamma Capricorni has a 2.40 au companion that is very likely a white dwarf.}
\end{figure*}

Gamma Capricorni ($\gamma$ Cap, HIP 106985, HD 206088, HR 8278, 40 Cap) is a naked-eye star (Figure \ref{fig:sky_image}) and with $V=+3.67$ the fourth brightest star in Capricornus. It bears the traditional (IAU-approved) proper name Nashira (``the lucky one'') of Arabic origin. It is a chemically peculiar Am star of somewhat puzzling nature \citep[see e.g. Appendix B.27 in][who also detected RV variations but could not conclude on their origin]{Carrier2002}. \cite{Abt1995} gave it a spectral type kF0hF1VmF2. A detailed abundance analysis in \cite{Erspamer2003} yielded remarkably high enrichment of s-process elements even compared to other Am stars. 

$\gamma$ Cap has several astrometric anomalies that point to the presence of a companion. These include significant proper motion changes between Gaia DR3 \citep{Gaia23} and Hipparcos \citep{Hipparcos1997} $\Delta v = 4.8 \pm 0.4 \text{ km}\text{ s}^{-1}$ as well as between Gaia DR3 and Gaia DR2 \citep{Gaia18} $\Delta v = 8.2 \pm 0.5 \text{ km}\text{ s}^{-1}$, wildly discrepant parallaxes (19 to 28 mas) between the different astrometric catalogs and large RUWE (6.6) and astrometric noise (3.15 mas) in Gaia DR3. In fact, $\gamma$ Cap was one of the 235 stars for which the original Hipparcos catalog reported an astrometric orbital solution. The orbit was assumed circular and had a period $P=800\pm75\text{ days}$, photocenter semi-major axis $a_{\mathrm{phot}} = 12.9\pm1.6 \text{ mas}$, inclination $i=68.1\pm6.0^{\circ}$ and longitude of the ascending node $\Omega=63.2\pm4.5^{\circ}$ (or $\Omega + 180^{\circ}$ due to the degeneracy of a pure astrometric solution). We note that an Gaia DR3 did not report an astrometric orbit for $\gamma$ Cap. 

In this paper we combine radial velocity, astrometric and interferometric measurements of $\gamma$ Cap to show that its companion is almost certainly a white dwarf. We also suggest that the primary is a hybrid Am barium star. This paper is organized as follows. In Section \ref{sec:data} we summarize the data. In Section \ref{sec:results} we solve for the orbital and stellar parameters of the system. In Section \ref{sec:discussion} we show that the companion is almost certainly a WD and discuss the relevance of the system as well as its past and future evolution. A brief conclusion can be found in Section \ref{sec:conclusion}. 

\section{Data} 
\label{sec:data}

\subsection{Previous radial velocities}

We used the Bibliography of stellar radial velocities \citep{Abt1972} to find references containing past radial velocity (RV) measurements of Nashira, which are collected in Table \ref{table:rv} for convenience. The data include eight observations from Lick Observatory collected between 1899 and 1919 \citep{Campbell1911,Campbell1928} and ten observations from Cape Observatory collected between 1908 and 1926 \citep{Jones1928}. We acknowledge the publicly available scans from HathiTrust which allowed us to get the data from the Lick Observatory Bulletins that do not have a source available in NASA ADS. 

In addition, there is a set of 24 RV measurements of Nashira reported in \cite{Carrier2002}. The observations were made with the CORAVEL scanner between 1979 and 1997. A single RV measurement (JD=2451741.795; $v_{\mathrm{z,A}}=-27.58 \text{ km}\text{ s}^{-1}$) with the CORALIE spectrograph was reported in \cite{Erspamer2003}. Finally, there is also a reported RV (JD=2451777.448, $v_{\mathrm{z,A}}=-29.57 \text{ km}\text{ s}^{-1}$) in the ELODIE archive \citep{Baranne1996,Moultaka2004}. 

\subsection{Spectra}

We complemented the RV data using two archival ``ready-to-use'' reduced spectra of $\gamma$ Cap. One spectrum was taken in 2013 with the HERMES spectrograph on the 1.2-meter \textit{Mercator} telescope and was downloaded from the MELCHIORS library \citep{Royer2024}. It has a wavelength coverage of 3760-9000{\AA} and an average spectral resolution $R=\frac{\lambda}{\Delta\lambda}=85,000$. Another spectrum was taken in 2014 with the Narval spectropolarimeter \citep{Auriere2003} and was downloaded from the PolarBase archive \citep{Petit2014}. It has a wavelength coverage 3700-10500{\AA} and a spectral resolution $R=65,000$.

Table \ref{table:spectra} provides further details of the spectral observations including dates and exposure times. Both spectra had already been corrected to the solar system barycentric frame. All the wavelengths are in air. 

\begin{table}
\centering
\caption{\label{table:spectra} Details of spectral observations of Gamma Capricorni.}
\begin{tabular}{cccc}
\hline \hline
\shortstack{date\\MJD} & Instrument & Exposure time & \shortstack{$v_{\mathrm{z,A}}$\\($\text{ km}\text{ s}^{-1}$)} \\ [0.3cm]

\shortstack{2013-06-06\\56449.209} & HERMES & 120 s & $-34.8\pm0.5$ \\ [0.3cm]

\shortstack{2014-09-11\\56911.899} & Narval & 380 s & $-33.4\pm0.5$ \\ [0.3cm]

\hline
\end{tabular}
\end{table}

\subsection{Hipparcos IAD}

We downloaded the Hipparcos \citep{Hipparcos1997} Intermediate Astrometric Data (IAD) for Nashira using the public python package \texttt{htof} \citep{Brandt2021}. The data consists of the 1D abscissae residuals from two independent consortia, namely NDAC (Northern Data Analysis Consortium) and FAST (Fundamental Astronomy by Space Techniques). 

There is a second set of abscissae values from the Hipparcos re-reduction \citep{vanLeeuwen2007}, which applied a more aggressive model for satellite attitude correction. This led to a reduction of the errors at the expense of possibly ``fitting out'' binary orbital motion \citep[see e.g.][for details]{Brandt2023}. We downloaded the corresponding abscissae from the 2014 Java tool provided by the ESA website. 

For both datasets we reconstructed the original abscissae values based on the abscissa residuals, the reported parallax and the reported proper motions. See \cite{Waisberg2026} for details. 

\subsection{VLTI/GRAVITY}

Nashira was observed with VLTI/GRAVITY \citep{GRAVITY17} on 2024-10-19 (MJD=60602.07) with the four 1.8-m Auxiliary Telescopes (ATs) in the configuration A0-G1-J2-K0, which provided a maximum projected baseline of 123.5 meters, corresponding to a maximum angular resolution of 3.7 mas in the working NIR K band. The observations were performed in single-field mode, wherein half of the photons were used to fringe track at low spectral resolution (R=22) and the other half were integrated in 10-seconds exposures in high spectral resolution (R=4,000). There were two consecutive observations containing 16 exposures each. The seeing during the observations decreased from 0.85 to 0.56" (@500 nm). The K3III star HD 205624 \citep[with angular diameter $\theta = 0.93\pm0.08 \text{ mas}$;][]{Bonneau2006,Chelli2016} was observed immediately after Nashira as interferometric calibrator. The data were reduced with the ESO GRAVITY pipeline v.1.6.6 \citep{Lapeyrere14}. Each file consists of of a set of squared visibilities for each of the six baselines and closure phases for each of the four triangles. We make use of the science (high spectral resolution) data for the scientific analysis. 

\section{Results}
\label{sec:results}

\subsection{New radial velocities}

The detailed spectral analysis of $\gamma$ Cap in \cite{Erspamer2003} yielded $T_{\mathrm{eff}} = 7100\pm150 \text{ K}$, $\log g = 3.45 \pm 0.15$, a microturbulent velocity $\xi = 2.5 \text{ km}\text{ s}^{-1}$ and $v \sin i = 42 \text{ km}\text{ s}^{-1}$. In order to measure the RVs for the two archival spectra, we used cross-correlation against a synthetic spectrum from the BOSZ spectral library \citep{Bohlin2017,Meszaros2024} based on a MARCS model atmosphere \citep{Gustafsson2008} with $T_{\mathrm{eff}} = 7100 \text{ K}$, $\log g = 3.5$, $\xi = 2 \text{ km}\text{ s}^{-1}$ and metallicity of +0.25 dex (in order to mimic the enhanced abundance of iron-peak elements, whose lines dominate the spectrum). We applied rotational and instrumental broadening to the model spectrum. The cross-correlation was performed within several wavelength regions rich in metallic lines and the scatter between the resulting RVs was used to estimate the uncertainty. The RVs are reported in Table \ref{table:rv}.

\subsection{Orbital parameters}

We started by using the set of previous and new RVs to search for a radial velocity curve solution. We constructed a three-dimensional grid in orbital period $P$ between 600 and 1000 days with a step of 0.1 day, eccentricity $e$ between 0 and 0.9 with a step of 0.05 and time of periastron $T_p$ between $-P/2$ and $+P/2$ with a step of 2 days. For each point in the grid, the remaining parameters (namely, argument of pericenter $\omega_A$, RV semi-amplitude $K_A$ and systemic velocity $\gamma$\footnote{At first we allowed for constant velocity offsets between the different datasets to account for possible instrumental effects, but found that the $\gamma$'s are statistically consistent within their errors.}) can be calculated fast by linear least squares. We found a significant global minimum at $P=840 \text{ days}$ and $e=0.1$, thereby confirming that $\gamma$ Cap is a single-lined spectroscopic binary (SB1) and that the Hipparcos astrometric binary solution is real. 

\begin{figure*}[]
\centering
\includegraphics[width=0.8\textwidth]{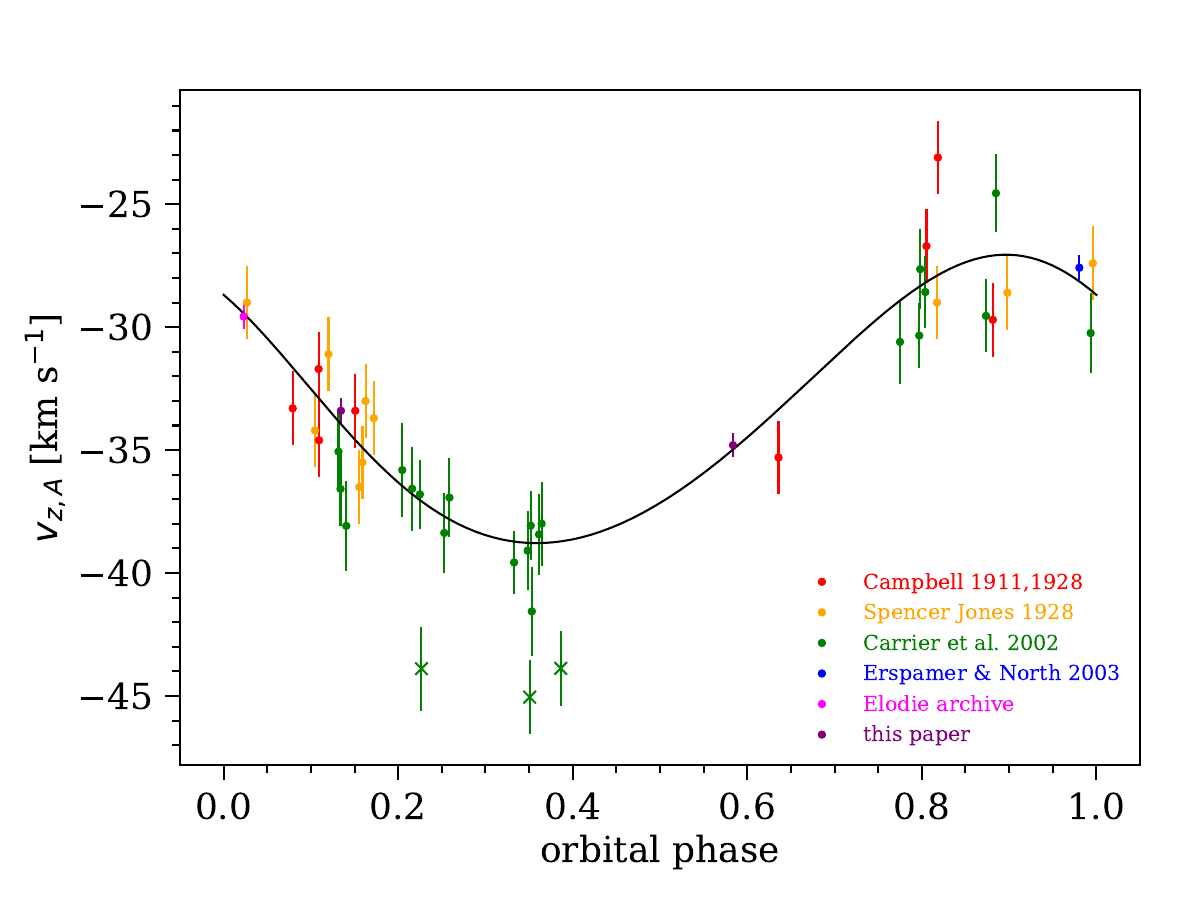} \\
\includegraphics[width=0.8\textwidth]{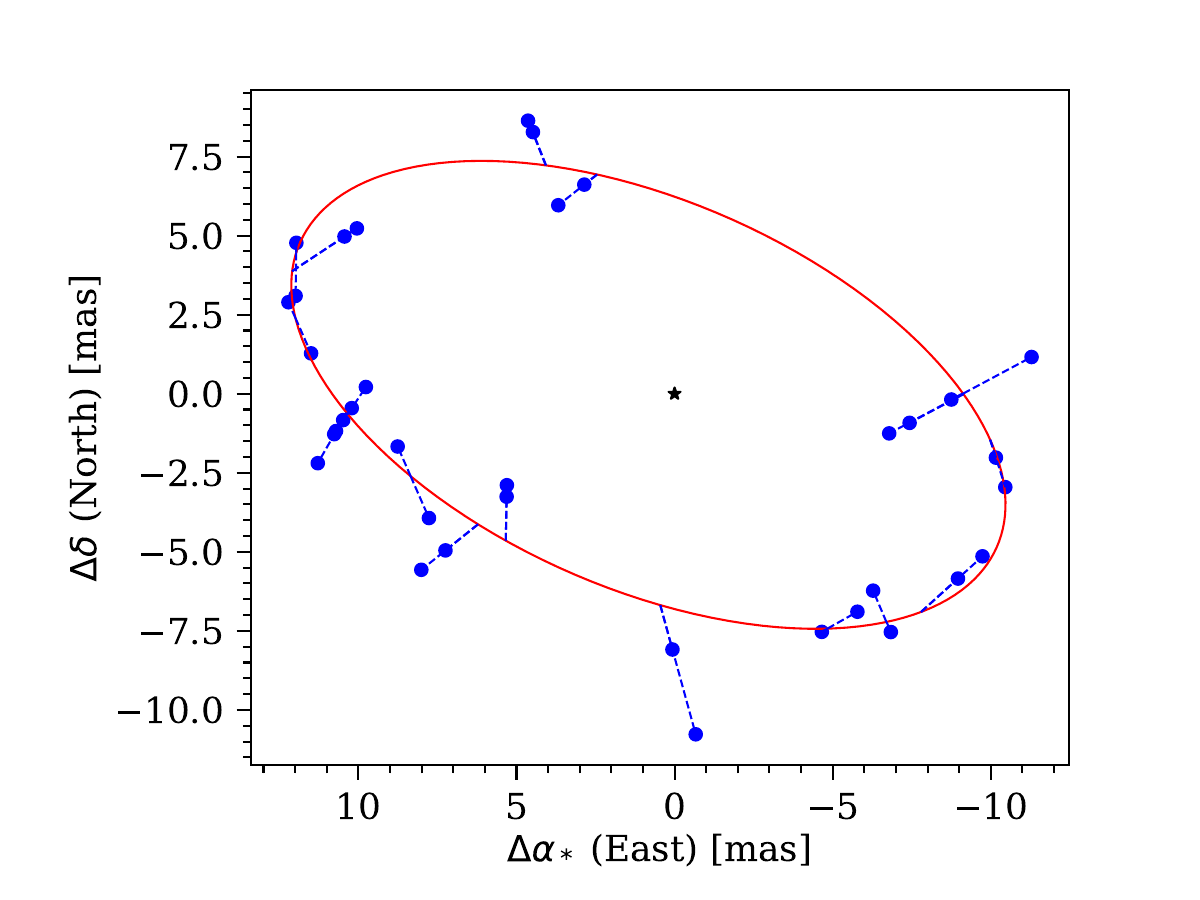}
\caption{\label{fig:orbital_fit} \textbf{Top:} Radial velocity data of the primary star of Gamma Capricorni (colored) and the corresponding best fit orbital solution (black line). The data span 115 years (1899-2014) or 50 orbits. \textbf{Bottom:} Abscissae residuals (blue) of the original 1997 Hipparcos Intermediate Astrometric Data (IAD) against the best fit photocenter orbit (red). Each abscissa is linked to its corresponding position on the orbit and the distance is given by the residual and plotted along the scan angle direction. The data span 2.7 yrs (1990.25-1992.95) or 1.2 orbits.}
\end{figure*}

We then performed a non-linear least squares fit to the RVs with all parameters allowed to vary and starting from the global minimum in the grid. This allowed us to estimate the errorbars for the data with unreported errors based on the residuals. We found that the \cite{Erspamer2003} and Elodie archive data points have an error of about 0.5 $\text{ km}\text{ s}^{-1}$ that is comparable to our two new data points, while the old Lick and Cape observations have an error of about 1.5 $\text{ km}\text{ s}^{-1}$ that is comparable with the reported errors in \cite{Carrier2002}. We also excluded from the fit the three data points in \cite{Carrier2002} with $v_z < -43.5 \text{ km}\text{ s}^{-1}$ since they are very clear outliers (marked with crosses in Figure \ref{fig:orbital_fit} (top)). 

In a next step we performed a joint non-linear least squares fit between the RVs and the Hipparcos IAD data. The latter share some parameters with the RV curve (namely, $P$, $e$, $\omega$ and $T_p$) and requires adding the remaining three Keplerian parameters, namely the orbital inclination $i$, the longitude of the ascending node $\Omega$ and the (angular) photocenter semi-major axis $\alpha_{\mathrm{phot}}$, as well as the parallax, the systemic (center-of-mass) proper motions and a fixed 2d-offset $(\Delta \alpha_*, \Delta \delta)$ from the reference position. In order to remove possible degeneracies, we fixed the proper motion to that calculated from the Gaia DR3-Hipparcos positional difference over 2016.0-1991.25=24.75 years, namely $(v_{\alpha_*},v_{\delta}) = (189.55, -22.98) \text{ mas}\text{ yr}^{-1}$. Because this interval is much longer than the orbital period, it allows to constrain the center-of-mass proper motion to very good accuracy and precision. 

When using the original Hipparcos 1997 IAD, we recover a solution consistent with the previously reported astrometric orbit in the original Hipparcos catalog but with smaller errors due to the addition of the radial velocity data. The NDAC and FAST data yield consistent solutions, so we used both of them for our final solution. The combined RV and astrometric solution has a $\chi^2 = 73$  for 67 degrees of freedom. Table \ref{table:orbital_fit} reports the corresponding orbital parameters. The errors were estimated by re-sampling the data according to the errors 5000 times and finding the 16th and 84th percentile thresholds of the resulting best fit parameter distributions ($1\sigma$). The full distribution plots for the orbital parameters can be found in Figure \ref{fig:orbital_fit_corner}. Figure \ref{fig:orbital_fit} (top) shows the radial velocity data (colored) and the best fit solution (solid black) as well as (bottom) the Hipparcos data (blue) and the best fit sky-projected photocenter orbital solution (solid black). We plot the latter in the typical way that is used to illustrate the Hipparcos one-dimensional abscissae \citep[e.g.][]{Torres2007,Sahlmann2011}, namely: each abscissa is connected to the corresponding point on the orbit at its given epoch through the dashed lines and the distance between them is given by the difference between the model and the measured abscissa projected along the scan angle direction. More details of the IAD residuals with and without including the photocenter orbit in the model can be found in Figure \ref{fig:hip_details}. 

\begin{table}
\centering
\caption{\label{table:orbital_fit} Orbital and stellar parameters for Gamma Capricorni.}
\begin{tabular}{ccc}
\hline \hline

& & source \\ 

\shortstack{$K_A$\\($\text{ km}\text{ s}^{-1}$)} & $5.9\pm0.3$ & RV \\ [0.3cm]

$e$ & $0.085_{-0.030}^{+0.056}$ & RV + astrometry \\ [0.3cm]

\shortstack{$\omega_A$\\(deg)} & $44\pm30$ & RV + astrometry \\ [0.3cm]

\shortstack{$P$\\(days)} & $840.22\pm0.25$ & RV + astrometry \\ [0.3cm]

\shortstack{$T_p$\\(MJD)} & $18148.8\pm67.5$ & RV + astrometry \\ [0.3cm]

\shortstack{$i$\\(deg)} & $59.3\pm5.2$ & astrometry \\ [0.3cm]

\shortstack{$\Omega$\\(deg)} & $246.3\pm3.5$ & astrometry \\ [0.3cm]

\shortstack{$\alpha_{\mathrm{phot}}$\\(mas)} & $12.0\pm1.0$ & astrometry \\ [0.3cm]

\hline \\ [0.1cm]

\shortstack{$\gamma$\\($\text{ km}\text{ s}^{-1}$)} & $-33.3\pm0.2$ & RV \\ [0.3cm]

\shortstack{plx(mas)\\d(pc)} & \shortstack{$24.1\pm0.8$\\$41.5\pm1.4$} & astrometry \\ [0.3cm]

\shortstack{$\Delta \alpha_*$\\(mas)} & $-1.0\pm1.0$ & astrometry \\ [0.3cm]

\shortstack{$\Delta \delta$\\(mas)} & $0.0\pm0.7$ & astrometry \\ [0.3cm]

\hline \\ [0.1cm]

\shortstack{$a_A \sin i$\\(au)} & $0.452 \pm 0.025$ & $K_A$, $P$, $e$ \\ [0.3cm]

\shortstack{$M_{A}$\\($M_{\odot}$)} & $2.06\pm0.08$ & stellar evolution \\ [0.3cm]

\shortstack{$M_{B}$\\($M_{\odot}$)} & \shortstack{$0.58\pm0.07$\\$0.53_{-0.05}^{+0.09}$\\$0.56\pm0.06$} & \shortstack{$K_A$,$P$,$e$,$i$,$M_A$\\$a_{\mathrm{phot}}$,$d$,$P$,$M_A$,$f\approx0$\\combined} \\ [0.3cm]

\shortstack{$a$\\(au)} & \shortstack{$2.40\pm0.02$\\$2.49\pm0.31$\\\\$2.36\pm0.32$} & \shortstack{$P$,$M_A$,$M_B$\\$a_A \sin i$,$i$,$M_A$,$M_B$\\$\alpha_{\mathrm{phot}}$,$d$,$M_A$,$M_B$} \\ [0.3cm]

\hline \\ [0.1cm]

\shortstack{$T_A$\\($K$)} & $7100\pm150$ & \cite{Erspamer2003} \\ [0.3cm]

\shortstack{$R_A$(mas)\\($R_{\odot}$)} & \shortstack{$0.47\pm0.04$\\$4.2\pm0.4$} & \shortstack{interferometry\\+ distance} \\ [0.3cm]

\shortstack{$\tau$\\(Myr)} & $\gtrsim 1000$ & stellar evolution \\ [0.3cm]

\shortstack{$T_B$\\($K$)} & $\lesssim 20,000$ & TD-1 photometry \\ [0.3cm]

\shortstack{$R_B$\\($R_{\odot}$)} & $\sim 0.013 R_{\odot}$ & \shortstack{WD mass-radius\\relation} \\ [0.3cm]

\hline \\ [0.1cm]

\hline
\end{tabular}
\tablenotetext{0}{Notes:}
\tablenotetext{0}{The uncertainties correspond to the 16\% and 84\% percentiles of the parameter distributions.}
\end{table}

Meanwhile, when using the Hipparcos 2007 re-reduction IAD, the best fit $\alpha_{\mathrm{phot}}$ collapses to zero and we could not find an acceptable fit for the astrometric data. This can be explained by the aggressive modelling of the Hipparcos satellite attitude that was applied in the 2007 re-reduction leading to overfitting that interferes with and suppresses the photocenter orbital motion. This is consistent with the raw IAD data having much smaller residuals relative to the 5-parameter fit in the 2007 data compared to the 1997 data. This issue has been previously discussed \citep[e.g.][]{Brandt2021,Brandt2023} and in fact we have recently run into a similar situation for another binary \citep{Waisberg2026}. $\gamma$ Cap therefore provides another warning case that the Hipparcos 2007 IAD should not be used to constrain or model photocenter orbital motion. 

\subsection{Angular radius and companion non-detection}
\label{subsec:gravity}

The VLTI/GRAVITY interferometric data is shown in Figure \ref{fig:gravity} (colored). There is no signature of a companion (in particular, the closure phases are consistent with zero). The squared visibilities show a smooth decay as a function of the magnitude of the projected baseline consistent with a circularly symmetric partially resolved disk. The best fit limb-darkened disk model is shown in solid black lines and yields a radius $R = 0.47 \pm 0.04 \text{ mas} \leftrightarrow 4.2 \pm 0.4 R_{\odot}$. We adopted a linear limb-darkening coefficient $u_{\lambda} = 0.19$ for a $\log g = 3.5$, $T_{\mathrm{eff}}=7250 \text{ K}$, solar metallicity star \citep{Claret2011}. Because the star is only partially resolved, limb-darkening has a very minor effect, increasing the stellar radius by only 1.4\% compared to a uniform disk. 


\begin{widetext}
\begin{center}
\includegraphics[width=0.95\textwidth]{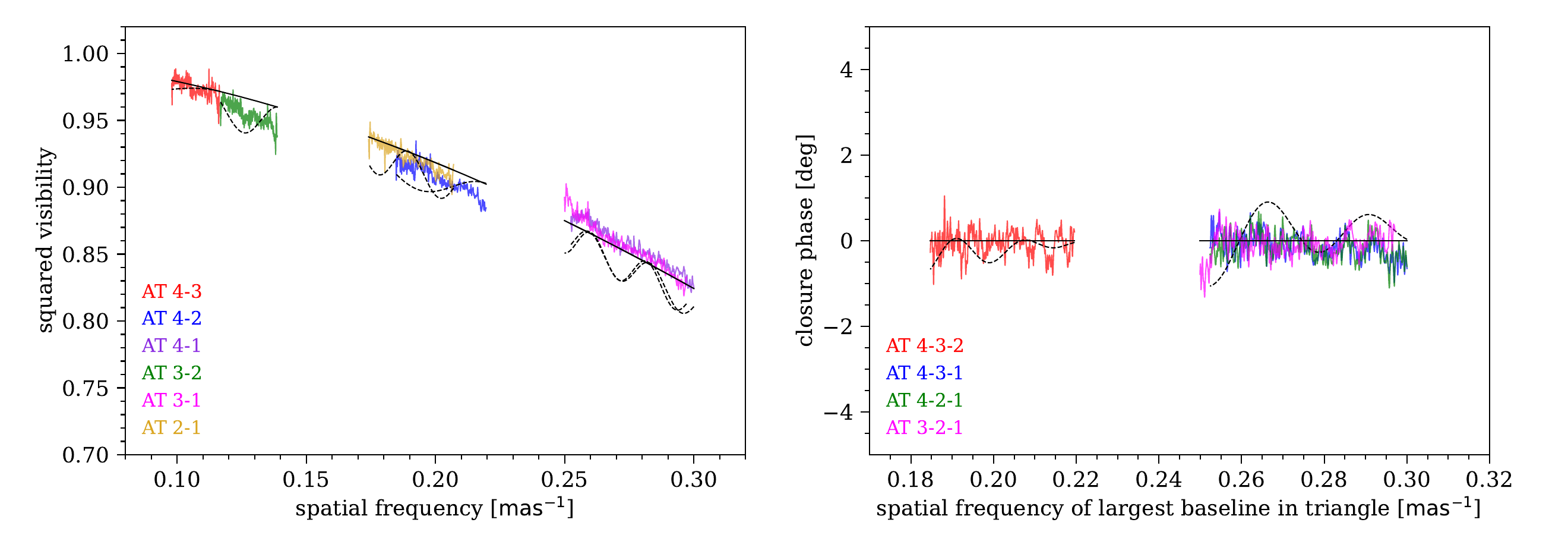}
\captionof{figure}{VLTI/GRAVITY data (colored) for Gamma Capricorni. The solid black lines show the best fit model for the resolved circular disk of the primary. The dashed black lines show the expected binary signature if there were a $0.56 M_{\odot}$ red dwarf companion at its expected position.}
\label{fig:gravity}
\end{center}
\end{widetext}

\begin{figure}[]
\centering
\includegraphics[width=\columnwidth]{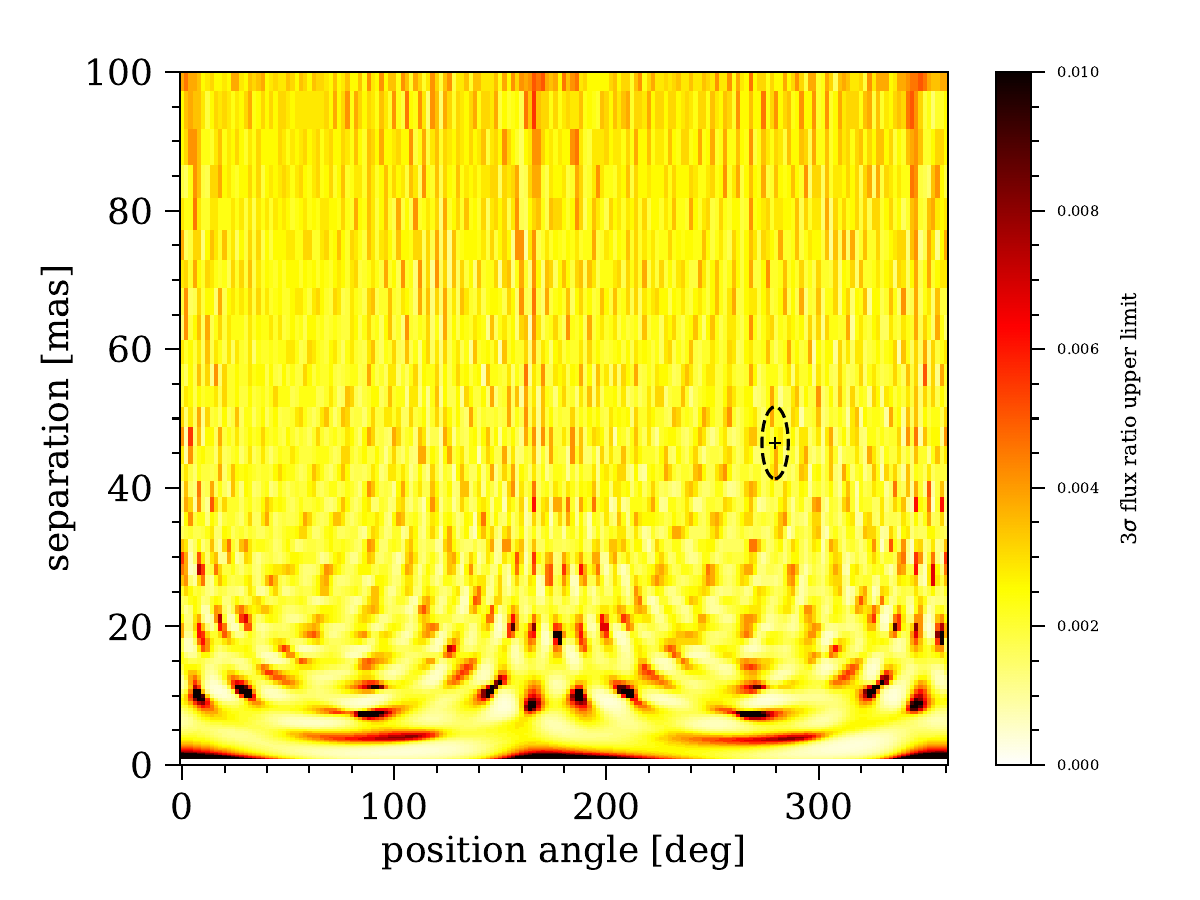}\\
\includegraphics[width=\columnwidth]{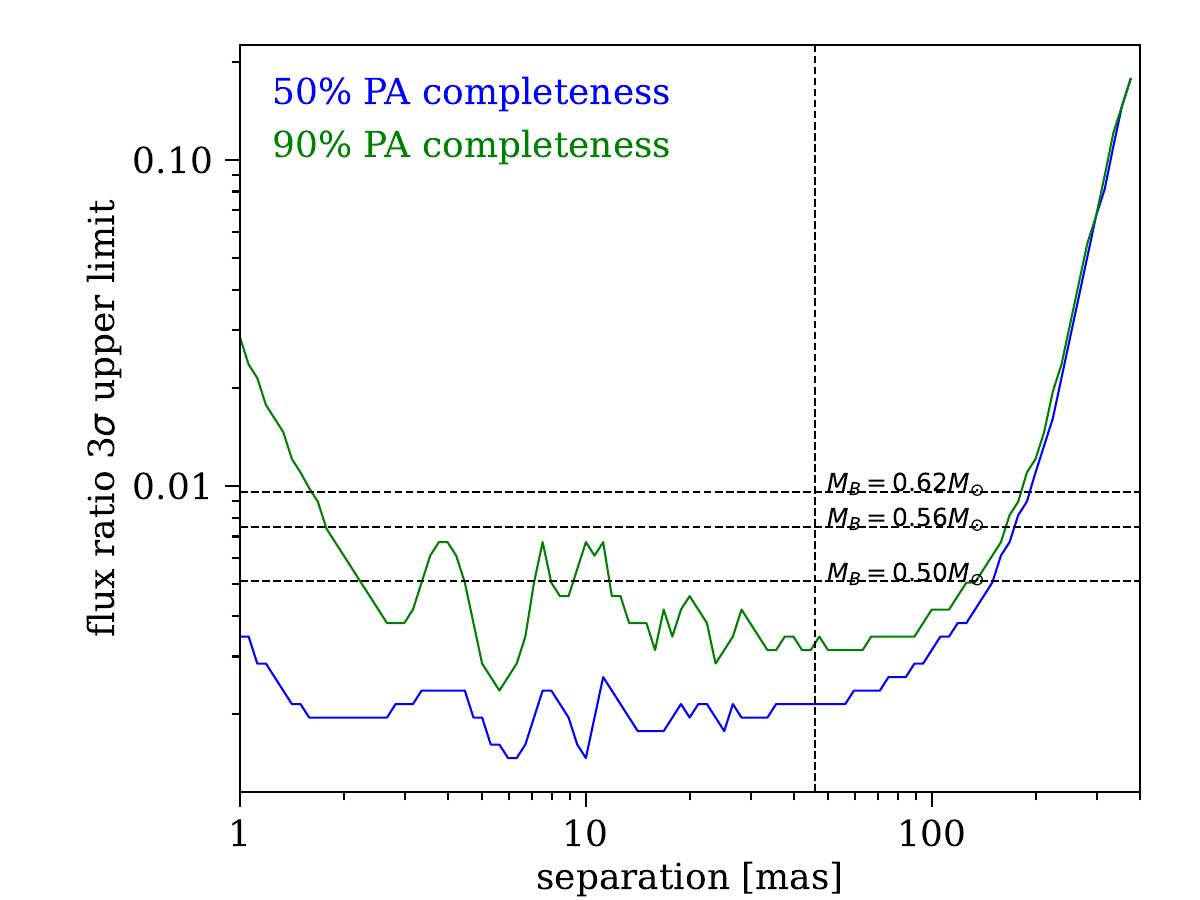}\\
\caption{\label{fig:gravity_upper_limit} \textbf{Top:} 3$\sigma$ upper limit on the K band flux ratio of a companion to $\gamma$ Cap based on the VLTI/GRAVITY data. The dashed ellipse shows the expected position of the companion at the VLTI/GRAVITY epoch based on the orbital parameters. \textbf{Bottom:} Same as above but as a function of separation only for different position angle completeness. The dashed vertical line marks the expected separation of the companion at the VLTI/GRAVITY epoch. The horizontal lines show the expected K band flux ratio for red dwarfs of different masses.}
\end{figure}

We follow the method described in \cite{Waisberg2023} to calculate the 3$\sigma$ K band flux ratio upper limit for a companion as a function of the 2D projected separation. Figure \ref{fig:gravity_upper_limit} shows the results in the top panel, while the bottom panel shows the same results as a function of separation only for 50\% and 90\% completeness in position angle. Note that the granular aspect in the 2D plot is real and caused by the very limited uv-coverage: there are positions where it is easier to ``hide'' a companion. 

\subsection{Primary evolutionary mass}
\label{subsec:primary_mass}

\cite{Erspamer2003} reported an effective temperature $T_{\mathrm{eff}}=7100\pm150 \text{ K}$ for $\gamma$ Cap based on spectral analysis. Our measured angular radius and distance yield a physical radius $R_A = 4.2 \pm 0.4 R_{\odot}$. We can therefore estimate the mass of the primary star from stellar evolutionary models\footnote{We note that $\log g = 3.45 \pm 0.15$ can only provide very loose bounds on the mass $1 M_{\odot} \lesssim M_A \lesssim 3 M_{\odot}$.}.  

\begin{figure}[]
\centering
\includegraphics[width=\columnwidth]{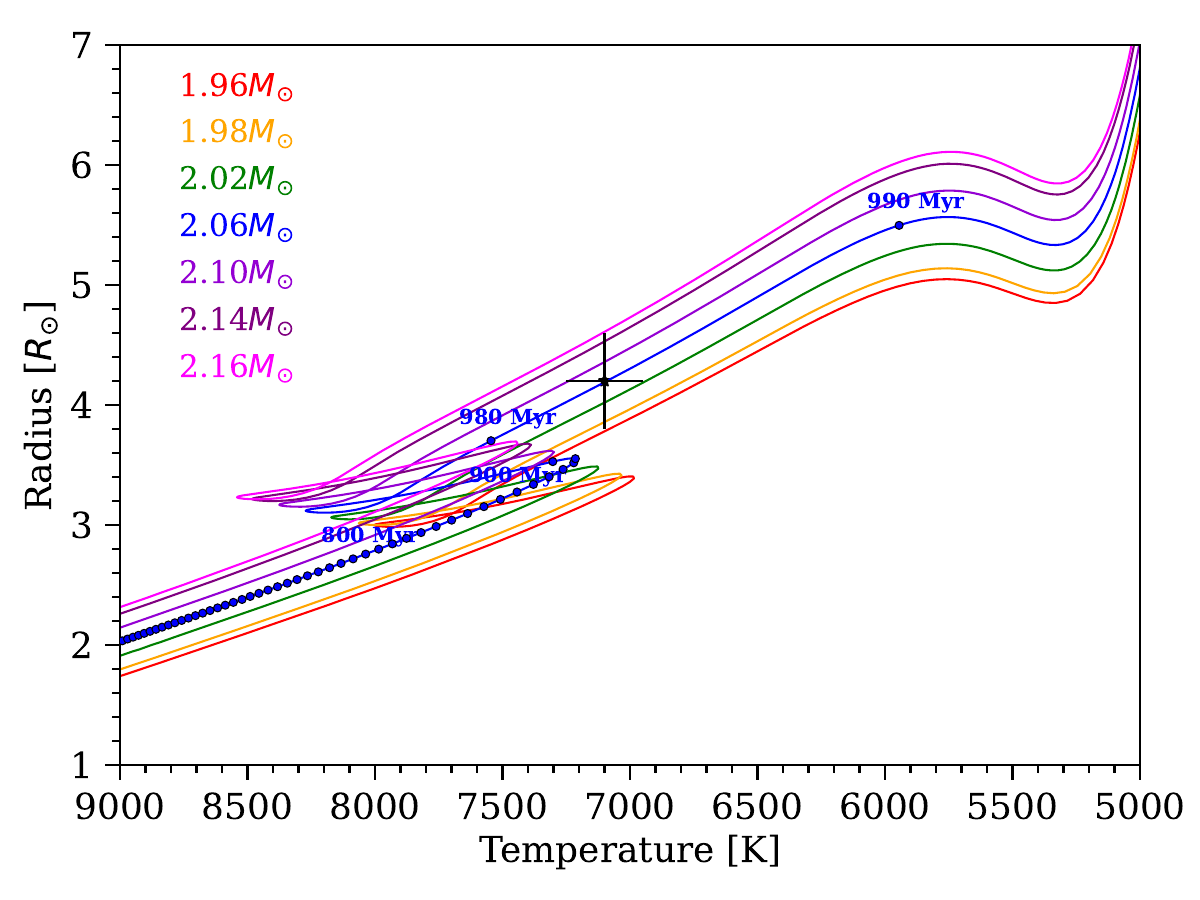}
\caption{\label{fig:evolutionary_tracks} \texttt{MIST} stellar evolutionary tracks for a range of different masses. The blue dots mark 10 Myr intervals. The measured parameters for Gamma Cap A are plotted in black.}
\end{figure}

In Figure \ref{fig:evolutionary_tracks} we plot evolutionary models from \texttt{MIST} \citep{Dotter16,Choi16,Paxton11,Paxton13,Paxton15} for a range of stellar masses and age $\tau > 100 \text{ Myr}$. The blue dots mark 10 Myr intervals on the $2.06 M_{\odot}$ track. The models have solar metallicity and zero rotational velocity. 

The position of $\gamma$ Cap A is plotted in black and imply that it is a 1 Gyr old $M_A=2.06\pm0.08 M_{\odot}$ star that has just become a subgiant (i.e. exhausted core hydrogen and started to expand). It is intriguing that according to the evolutionary tracks this is a very fast (lasting less than 20 Myr) and therefore unlikely phase to observe such a star. We note that using the evolutionary tracks for a rotational velocity of 0.4 times critical instead leads to a negligible change in the stellar parameters. 

\section{Discussion}
\label{sec:discussion}

\subsection{The mass of the companion}

There are two ways to calculate the mass of the companion. In both of them we need to make use of the primary's evolutionary mass $M_A=2.06\pm0.08 M_{\odot}$. 

The first one is to use the RV semi-amplitude of the primary together with the orbital inclination from the photocenter astrometric orbit: 

\begin{align}
& \frac{M_B^3}{(M_A+M_B)^2} = \frac{P}{2\pi G} \frac{K_A^3 (1-e^2)^{3/2}}{\sin^3 i} = 0.029\pm0.008 M_{\odot} \\ & \Rightarrow M_B = 0.58\pm0.07 M_{\odot}
\end{align}

Alternatively, we can use the photocenter semi-major axis 

\begin{align}
a_{\mathrm{phot}} = a \left ( \frac{q}{1+q} - \frac{f}{1+f} \right )
\end{align}

\noindent where $q = \frac{M_B}{M_A}$ is the mass ratio and $f=\frac{f_B}{f_A}$ is the flux ratio in the Hipparcos band, combined with Kepler's Third Law: 

\begin{align}
\frac{(1+q)^2}{q^3} \frac{(1+f)^3}{(1-\frac{f}{q})^3} = \frac{GM_A}{a_{\mathrm{phot}}^3} \left ( \frac{P}{2\pi} \right )^2 = 90 \pm 30
\end{align}

Based on the non-detection of the companion in the VLTI/GRAVITY data, we approximate $f \simeq 0$ so that 

\begin{align}
q = 0.260_{-0.027}^{+0.045} \Rightarrow M_B = 0.53_{-0.05}^{+0.09} M_{\odot}
\end{align}

The fact that the two estimates for $M_B$ are consistent is reassuring. Alternatively, consistency can also be checked by noting that for $f\simeq0$ the photocenter semi-major axis is the same as that of the primary so that 

\begin{align}
K_A = \frac{2 \pi}{P} \frac{a_{\mathrm{phot}}\sin i}{(1-e^2)^{1/2}}
\end{align}

\noindent Using the measured $a_{\mathrm{phot}}$ we predict $K_A=5.6\pm0.8 \text{ km}\text{ s}^{-1}$, which is fully consistent with the measured value. 

Combining both estimates we adopt a mass $M_B = 0.56 \pm 0.06 M_{\odot}$ for the secondary. 

There are a handful of ways of calculating the binary semi-major axis based on the orbital parameters, masses and parallax; they are listed in Table \ref{table:orbital_fit}. By far the most precise is to use the period and the estimated masses and this yields $a = 2.40 \pm 0.02 \text{ au}$.

\subsection{The nature of the companion}

From its mass $M_B = 0.56 \pm 0.06 M_{\odot}$, the companion could be either a red dwarf or alternatively a white dwarf. A $0.50$,$0.56$,$0.62 M_{\odot}$ main-sequence star is expected to have an absolute $K_s$ band magnitude of $M_{K_s} = 5.64$,$5.22$,$4.95$ \citep{Pecaut2013}\footnote{$M_{K_s}$ is virtually independent of age for such a red dwarf after it has settled into the main-sequence in about 100-200 Myr of being born.}. Nashira has a Johnson K band magnitude $K_J = 3.04$ \citep{Ducati2002}, which corresponds to a 2MASS magnitude $K_s \simeq K_J - 0.044 \simeq 3.00$ \citep{Carpenter2001}. Its absolute magnitude is therefore $M_{K_s}\simeq-0.09$. The K band flux ratios for such red dwarf companions relative to the primary would then be $0.51\%$,$0.75\%$,$0.96\%$. In contrast, a white dwarf would have a much smaller K band flux by orders of magnitude. For a WD with mass $M=0.6 M_{\odot}$, the absolute $K_s$ band magnitude is $M_{K_s} \sim 11$ for $T_{\mathrm{eff}}=20,000 \text{ K}$ and $M_{K_s} \sim 13$ for $T_{\mathrm{eff}}=5,000 \text{ K}$ \citep{Bedard2020,Holberg2006,Tremblay2011}. In this case, the K band flux ratio relative to the primary would be $4 \times 10^{-3} \%$ and $6 \times 10^{-4} \%$, respectively. 

We can find the nature of the companion from the VLTI/GRAVITY observation. The orbital semi-major axis $a=2.40 \text{ au} \leftrightarrow 58 \text{ mas}$ falls within the optimal range for NIR interferometry. In fact, based on the full distribution of the orbital parameters we can calculate the projected separation vector of the companion relative to the primary at the VLTI/GRAVITY epoch $(\Delta \alpha_*, \Delta \delta) = (-45.5\pm5.2,+7.6\pm5.0) \text{ mas}$. This region is highlighted in Figure \ref{fig:gravity_upper_limit} (top). Within it, the $3\sigma$ flux ratio upper limit (\ref{subsec:gravity}) has a median of $0.20\%$ with a minimum of $0.12\%$ and a maximum of $0.38\%$. This is enough to exclude main-sequence red dwarfs with $M_{K_s} \lesssim 5.96$ or $M \gtrsim 0.45 M_{\odot}$. For illustration, the dashed lines in Figure \ref{fig:gravity} show the interferometric signal that would have been created by a $M_B=0.56 M_{\odot}$ red dwarf companion at its expected position, which are clearly inconsistent with the data. Figure \ref{fig:gravity_upper_limit} (bottom) shows that the expected flux ratios for red dwarfs with $M_B \gtrsim 0.50 M_{\odot}$ are safely above the upper limit at the expected separation $\rho=46 \text{ mas}$ of the VLTI/GRAVITY epoch. We conclude that the companion is most likely a white dwarf. 

It should be noted that there is a third option which is possible in principle albeit much less likely, namely that the companion is itself a pair of red dwarfs. For such a triple system to be stable, the pair would have to be very tight ($a\lesssim 0.1 \text{ au}$) so can be approximately treated as a single source but with a smaller flux. Due to the steep mass-luminosity relation, the most efficient way to reach a given mass with the lowest possible luminosity is to have an equal mass binary. For instance, a $0.28 M_{\odot} + 0.28 M_{\odot}$ M dwarf binary has a combined absolute magnitude $M_{K_s} \simeq 6.35$ \citep{Pecaut2013} and thereby a K band flux ratio of about $0.27\%$ relative to $\gamma$ Cap A. Given our flux ratio upper limits detailed above, we can exclude such a pair up to 3$\sigma$ within most of the expected position ellipse ($0.12-0.38\%$ with a median of $0.20\%$) but not all of it. 


Even so, in the case of Nashira there are additional hints for the companion being a WD such as the low eccentricity $e \lesssim 0.1$ (naturally explained by post-AGB tidal circularization) and the s-process and carbon abundances of the primary (see \ref{subsec:abundances}). In fact, unseen white dwarfs to intermediate-mass stars are believed to be rather common and may account for $\sim 20\%$ of companions to intermediate-mass stars with periods $P \sim 500\text{–}1000\text{ days}$ \citep{Murphy2018}. Therefore, we argue that considering all the evidence $\gamma$ Cap B is almost certainly a white dwarf. 

\subsection{Lack of X-rays and maximum white dwarf temperature}

Nashira was not detected by the ROSAT X-ray survey. A 1 deg search around it shows that the faintest sources detected have a count rate just above 0.02 ct$\text{ s}^{-1}$ \citep{Voges2000a,Voges2000b}; therefore, we can adopt a conservative upper limit of 0.02 ct$\text{ s}^{-1}$ for the 0.1-2.4 keV energy band of ROSAT\footnote{eROSITA can provide more stringent upper limits but $\gamma$ Cap is not in the eROSITA-DE sky region so that an upper limit is not readily available.}. We use the ROSAT count rate to flux conversion formula in \cite{Huensch1998} and a distance $d=41.5\text{ pc}$. 

For a hardness ratio $\text{HR1}\simeq-1$ appropriate for photospheric soft X-ray emission from a white dwarf, the resulting X-ray luminosity upper limit is $L_X < 1.3 \times 10^{28} \text{ erg}\text{ s}^{-1}$. This can be compared with the count rate 12.7 ct$\text{ s}^{-1}$ for Sirius (which at Nashira's distance would be 0.05 ct$\text{ s}^{-1}$), which corresponds to an X-ray luminosity $L_X = 3.1 \times 10^{28} \text{ erg}\text{ s}^{-1}$. Since the X-ray luminosity of Sirius is dominated by Sirius B, we can conclude that Nashira does not contain a WD as hot as Sirius B ($T \simeq 25 \text{ kK}$) although cooler WDs are not excluded by the X-ray upper limit.

For a hardness ratio $-0.5 \lesssim \text{HR1} \lesssim 0.2$ typical of coronal emission from a red dwarf, the resulting X-ray luminosity upper limit is $L_X < 2.4-4 \times 10^{28} \text{ erg}\text{ s}^{-1}$. This is right in the range of the expected coronal X-ray luminosity $L_X \sim 4 \times 10^{28} \text{ erg}\text{ s}^{-1}$ of 1 Gyr-old K/M dwarfs \citep[e.g.][]{Garces2011}. Therefore, the lack of an X-ray detection is consistent with the companion \textit{not} being a red dwarf (or a pair of M dwarfs).

\subsection{Constraints on the white dwarf temperature}

Ultraviolet data is particularly useful to detect and/or constrain the temperature of WD companions to main-sequence and giant stars. For Nashira, we can make use of the TD-1 photometry (in particular, its shortest wavelength filter at 1565{\AA}) to constrain the maximum temperature that a WD companion could have. Unfortunately, there are no IUE SWP or ultraviolet HST data for this target. 

\begin{figure}[]
\centering
\includegraphics[width=\columnwidth]{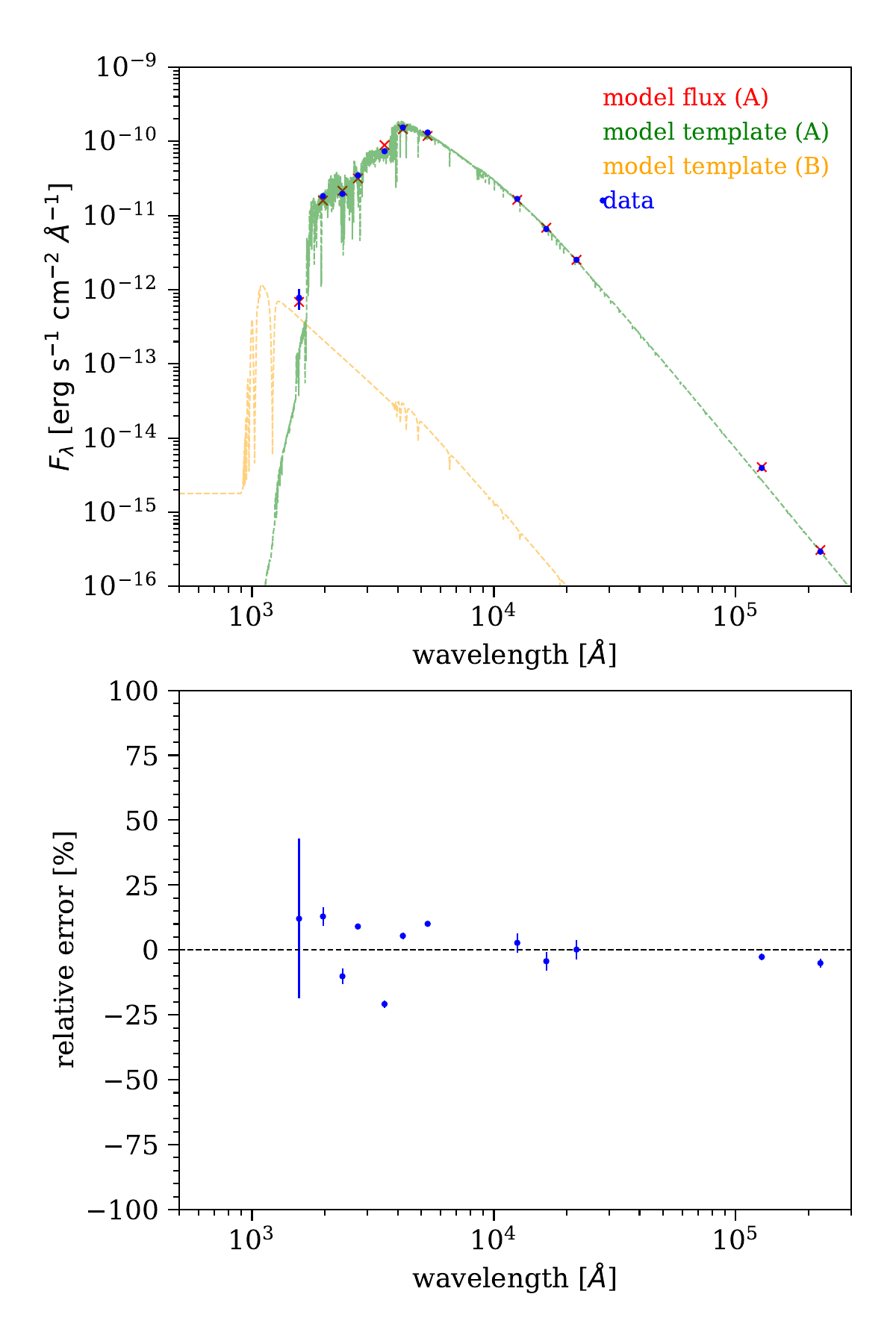}
\caption{\label{fig:photometry_fit} \textbf{Top:} Photometric data of Gamma Cap (blue) with the corresponding model fluxes (red) of the BOSZ spectral template for the primary (green). The yellow line shows a model template for the hottest white dwarf companion that could still be consistent with the TD-1 1565{\AA} photometry. \textbf{Bottom:} Corresponding residuals.}
\end{figure}

Figure \ref{fig:photometry_fit} plots space-based photometry including the four ultraviolet bands from TD-1 \citep{Thompson1978}, $V_T$ and $B_T$ from Tycho2 \citep{Hog2000} and W3 and W4 from WISE \citep{Wright2010}, as well as ground-based photometry including Johnson U, J, H and K bands \citep{Johnson1966,Ducati2002}. For the latter, we adopted conservative errors of 0.015 mag for U and 0.04 mag for the NIR bands. We note that (i) 2MASS \citep{Skrutskie2006} measurements have significantly larger errors for such a bright star; (ii) W1 and W2 were not used due to positive saturation flags; (iii) we refrained from using Gaia photometry for such a bright star also due to saturation issues. 

A BOSZ template with $T_{\mathrm{eff}} = 7100 \text{ K}$, $\log g =3.5$, metallicity of +0.25 dex and scaled to a radius $R = 4.2 R_{\odot}$ is plotted in green, with the resulting model photometry in red. The model fluxes over each filter were calculated using the public module \texttt{pyphot} \citep{Fouesneau20215}. Filter profiles for the TD-1 bands (not included in \texttt{pyphot}) were downloaded from the SVO Filter Profile Service \citep{Rodrigo2024}.

The photometric data is consistent with $\gamma$ Cap A, which confirms that $T_{\mathrm{eff,A}} = 7100 \text{ K}$ and $R_A = 4.2 R_{\odot}$ are accurate. We note that the UV fluxes are somewhat sensitive to the abundances due to strong line blanketing; creating a detailed spectral model taking into account the individual abundances is beyond the scope of this paper. 

In order to constrain the maximum temperature of the white dwarf, we make use of WD spectral templates from \cite{Koester2010}. We adopt the DA models with $\log g = 8.0$ and use a radius $R = 0.013 R_{\odot}$ appropriate for a mass $M_B \simeq 0.56 M_{\odot}$. We then find the maximum temperature allowed for the WD so that the net TD-1 1565{\AA} flux is still consistent with the measured one. We find that $T_{\mathrm{eff,B}} \lesssim 20,000 \text{ K}$. The corresponding template for this maximum temperature is plotted (yellow) in Figure \ref{fig:photometry_fit}. 

The cooling time for a $0.6 M_{\odot}$ WD to reach $T_{\mathrm{eff,B}} = 20,000 \text{ K}$ is about 60 Myr \citep{Bedard2020}, so that a cool WD is completely consistent with the relatively old $\tau \sim 1 \text{ Gyr}$ age of the primary. A more stringent temperature upper limit could possibly be obtained by extreme ultraviolet observations with the Hubble Space Telescope. However, it should be noted that in 1 Gyr the WD can cool to $T_{\mathrm{eff,B}} \sim 8,000 \text{ K}$, in which case a direct detection of the WD may be impossible. 

\subsection{One of the brightest stars with a white dwarf companion}

With the white dwarf nature of its companion firmly established, Nashira joins a select list of bright stars ($V\lesssim4$) with white dwarf companions. There are only three stars brighter than Nashira with a confirmed WD companion \citep{Holberg2013}, namely Sirius ($V=-1.47$) \citep{Bessel1844,Bond1862,Adams1915}, Procyon ($V=+0.39$) \citep{Bessel1844,Schaeberle1896} and Tianjinnan=Zeta Cygni ($V=+3.2$) \citep{Dominy1983,Barstow2001}\footnote{The stripped subdwarf companion to Regulus=Alpha Leonis is sometimes referred as a white dwarf but strictly speaking it is pre-white dwarf \citep{Gies2020}.}. We note that Mira, which also has a confirmed WD companion \citep{Sokoloski2010}, is brighter than Nashira at its lightcurve peak ($V\approx3$) but not on average. 

In all these cases except for Nashira, the orbital periods are very long of the order of decades to centuries and the WD has been spatially resolved. Zeta Capricorni ($V=+3.75$) -- the prototype barium star -- and Theta Hydrae ($V=+3.88$), both of which have an unresolved and hot WD companion detected in the extreme ultraviolet \citep{Boehm-Vitense1980,Burleigh1999}, complete the list of such $V\lesssim4$ stars. It should be noted that there is a fair number of bright giant stars that are suspected of having a WD companion due to their barium enrichment, namely Gamma Crucis ($V=+1.63$), Alpha Hydrae ($V=+2.00$), Delta Sagittarii ($V=+2.70$), Beta Bo\"otis ($V=+3.49$) and Eta Eridani ($V=+3.89$). However, there is so far no confirmation of a WD companion in any of these cases. 

\subsection{A hybrid Am and barium star?}
\label{subsec:abundances}

\begin{figure*}[]
\includegraphics[width=2.2\columnwidth]{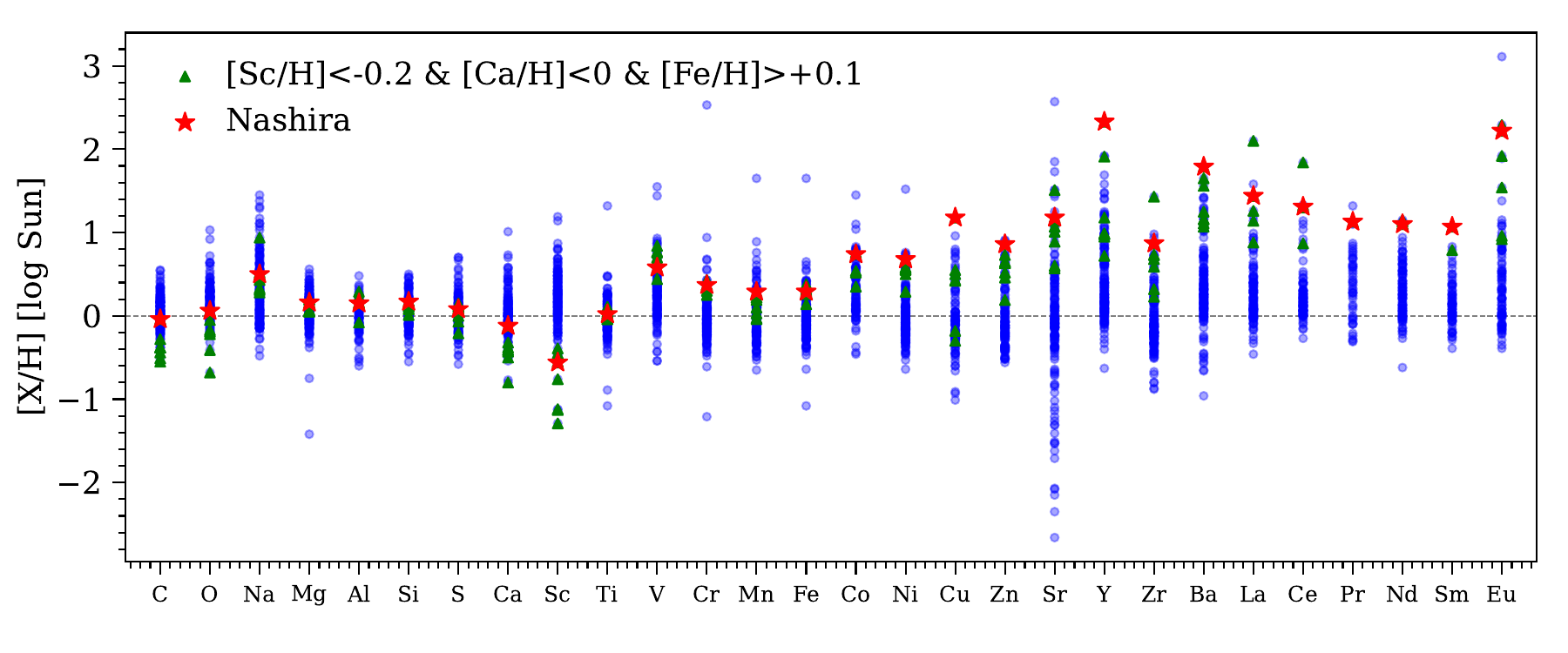}
\caption{\label{fig:abundances_all} Calculated abundances from \cite{Erspamer2003} for their sample of 140 A-F stars. Nashira stands out for its extreme s-process enrichment even among other Am stars (green) as well as for its solar carbon abundance.}
\end{figure*}

Figure \ref{fig:abundances_all} plots the abundances of the 140 A-F-type stars in \cite{Erspamer2003}. The sample includes normal as well as Am/Fm, Ap and metallic giants. Nashira stands out for having the highest enrichment among all stars in the s-process elements yttrium (2.33 dex above solar), barium (1.79 dex) and samarium (1.07 dex), as well as among the few highest in zirconium (0.87 dex), lanthanum (1.44 dex), cerium (1.31 dex), praseodymium (1.13 dex) and neodymium (1.10 dex). As in the case of traditional barium stars, the white dwarf companion can explain the extreme s-process enhancement through accretion of s-process enriched material when the former primary was in the thermally pulsing AGB stage \citep{Bidelman1951,McClure1980,Boffin1988,McClure1990,Rekhi2024}.

On the other hand, Nashira also shows abundance patterns typical of an Am star such as moderate calcium (-0.12 dex) and scandium (-0.56 dex) depletion and moderate iron-peak (e.g. Fe, Cr, Ni) enhancement. Therefore, the abundance pattern must have also been modified by radiative levitation/sinking in the relatively slow rotating atmosphere \citep[i.e. the typical mechanism operating in Am stars;][]{Abt1961,Michaud1970,Watson1970,Watson1971,Richer2000}. This can also explain additional abundance patterns such as a solar abundance of carbon (which would be significantly enhanced in a pure barium star but is often depleted in a regular Am star) and significant enrichment of europium (a dominantly r-process element) that cannot be explained by the mass transfer from a past AGB companion alone. 

The green triangles in Figure \ref{fig:abundances_all} show Am stars with [Sc/H]$<-0.2$, [Ca/H]$<0$ and [Fe/H]$>+0.1$. Although Am stars also show significant enhancement in s-process abundances due to radiative levitation, Nashira clearly stands out even among Am stars. Another peculiarity is its solar carbon abundance compared to carbon depletion in typical Am stars, which could be explained by the counteracting carbon enrichment from accretion of AGB material. We therefore suggest Nashira to be a hybrid Am and barium star. If true, this ties into the known problem of the lack of A-type barium stars: while giant barium stars span masses up to about $4 M_{\odot}$ \citep{Jorissen2019}, there are very few A-type ($M \gtrsim 1.6 M_{\odot}$) barium stars \citep[e.g.][]{Escorza2019} (which would be the natural progenitors for the heavier barium giants), with only a handful of candidates such as Beta Crateris and possibly IK Pegasi \citep{Smalley1996,Smalley1997,Hallakoun2026}. It is possible and indeed likely that some A-type barium stars are hiding among the Am class but have not been identified as such because of conflation between the two phenomena \citep[e.g.][]{Hakkila1989,Escorza2019,North2020}. We suggest that Am stars with standout s-process enrichment like Nashira are strong candidates for being hybrid Am and barium stars, which can be tested by discovering close white dwarf companions. 

\subsection{Progenitor and past evolution}

At least a moderate amount of mass $\Delta M \sim 0.1 M_{\odot}$ should have been transferred from the WD progenitor while it was an AGB star to its companion in order to create the s-process enrichment \citep[e.g.][]{Boffin1988}. Given the current semi-major axis $a=2.40 \text{ au} \leftrightarrow 520 R_{\odot}$, the fact that the binary should have expanded due to mass loss from the AGB progenitor and the fact that a $2-3 M_{\odot}$ AGB star can reach a radius $R \sim 500 R_{\odot}$, the latter likely filled its Roche lobe. 

Further evidence that some of the envelope of AGB star was stripped by its companion comes from the relatively low mass of the WD. Although there is significant scatter in the Initial-Final Mass Relation (IFMR) for WDs \citep[e.g.][]{Cummings2018,ElBadry2018,Marigo2020,Cunningham2024,Hollands2024}, it would be difficult to create a $0.56\pm0.06 M_{\odot}$ WD in 1 Gyr without some envelope stripping. Even for a WD mass of $0.62 M_{\odot}$, the various IFMRs imply a progenitor mass of $1.6-1.9 M_{\odot}$, which are less massive than $\gamma$ Cap A and would take 1.5-2.4 Gyr to become a WD. A $2.5 M_{\odot}$ star becomes a WD in 800 Myr and the latter cools to $T_{\mathrm{eff}} \lesssim 20,000 \text{ K}$ in about 60 Myr, which would be just enough to fit in the estimated 1 Gyr age of $\gamma$ Cap A. However, the expected WD mass in this case (without any envelope stripping) would be $0.68-0.69 M_{\odot}$ for the various IFMRs. 

We suggest instead that $\gamma$ Cap B most likely originated from a $M \sim 2.5 M_{\odot}$ progenitor that filled its Roche lobe at the late AGB stage, with the resulting mass transfer halting the growth of its CO core but only mildly (the late thermal pulses are needed for the s-process production and dredge-up). Higher progenitor masses for the WD are disfavored for several reasons: they would result in heavier white dwarfs even with some envelope stripping, heavy AGB stars are much less efficient in producing heavy s-process elements such as Ba \citep[e.g.][]{Karakas2016,Rekhi2024} and too heavy a donor star would tend to make the mass transfer process dynamically unstable \citep[e.g.][]{Hjellming1987}. In this scenario, $\gamma$ Cap A may have accreted a somewhat significant amount of mass $\Delta M \sim 0.5 M_{\odot}$ and rejuvenated. In this case, our estimated age of 1 Gyr assuming single star evolution should be taken as a lower limit to the age of the system. Even with rejuvenation the evolutionary mass $M_A$ should still be roughly accurate since convective core overshooting should adjust the core mass and interior structure to match the new total mass \citep[e.g.][]{Hellings1983,Braun1995}. Still, the fact that the primary appears to be in a very special (short duration) phase of its evolution (namely, it has just begun crossing the Hertzsprung Gap) is intriguing. 

\subsection{Future evolution}

As $\gamma$ Cap A evolves and becomes an AGB star itself, the system will likely become a D-type (dusty) symbiotic star. It will be first powered by wind accretion and subsequently by Roche lobe overflow once $\gamma$ Cap A fills its Roche lobe $R_{\mathrm{L,A}} \approx 0.49 a \simeq 250 R_{\odot}$ \citep{Eggleton1983} during thermal pulses (He shell flashes). Because the envelope is convective (thereby expanding upon mass loss) and the mass ratio $\frac{M_B}{M_A} \sim 0.3$ is low, mass transfer will likely be dynamically unstable \citep[e.g.][]{Hjellming1987} and lead to the formation of a common envelope. Because the orbit is wide and the envelope relatively loosely bound, there should be plenty of orbital energy to expel it. $\gamma$ Cap will therefore most likely become a close sub-Chandrasekhar WD binary. 

\section{Conclusion}
\label{sec:conclusion}

Nashira=Gamma Capricorni is a chemically peculiar Am star that has standout s-process enrichment and was previously reported as a wide $P \sim 800 \text{ days}$ astrometric binary in the original Hipparcos catalog. We combined historical, archival and new radial velocities spanning 115 years to show that it is indeed a $P=840 \text{ days}$ spectroscopic binary, confirming that the Hipparcos solution is real. A joint fit with the Hipparcos IAD provided a consistent solution for all the orbital parameters. It is worth noting that the IAD from the Hipparcos 2007 re-reduction appear to be significantly affected by the aggressive satellite attitude modeling and do not recover the expected photocenter orbital motion. 

Using the evolutionary mass of the primary star $M_A = 2.06 \pm 0.08 M_{\odot}$, the orbital solution imply a mass $M_B = 0.56 \pm 0.06 M_{\odot}$ for the companion. A VLTI/GRAVITY observation that did not detect the companion down to a K band flux ratio of about $0.2\%$ allows to exclude ($> 3 \sigma$) the possibility that it is a main-sequence red dwarf. A tight pair of less massive M dwarfs can be excluded at $3 \sigma$ throughout most of the expected position ellipse but not entirely. Further interferometric observations would certainly be useful to completely exclude this possibility. Just the same, considering the additional evidence for a WD companion such as the very low eccentricity, the s-process and carbon abundances and the lack of X-ray emission, we argue that the companion is almost certainly a white dwarf. To our knowledge this is the first time that a WD was ``discovered'' by combining a radial velocity/astrometric solution with a non-detection by optical/NIR interferometry. $\gamma$ Cap is the fourth visually brightest star for which a white dwarf companion has been discovered after Sirius, Procyon and Tianjinnan (Mira can be visually brighter than Nashira but only at its peak). 

Given the semi-major axis $a=2.40 \text{ au} \leftrightarrow 58 \text{ mas}$ and the extreme contrast ratio ($\Delta V \sim \Delta K \sim 12$), the WD cannot be directly detected in the optical or near-infrared bands. Ultraviolet photometry from the TD-1 $1565{\AA}$ band implies that the WD cannot be hotter than about 20 kK. Unfortunately, there are no archival IUE SWP observations of $\gamma$ Cap. The only hope of a direct detection of the WD would be with HST and would require the WD to be hotter than about 10 kK. Moreover, Nashira was not detected by ROSAT and more stringent X-ray upper limits could be provided by eROSITA-RU. 

We argue that despite $\gamma$ Cap clearly being an Am star its extreme s-process enrichment (including the highest Ba and Y abundance in the sample of 140 A-F stars in \cite{Erspamer2003}) and solar (rather than depleted) carbon abundance suggest that it is also a barium star (i.e. it accreted mass from a former AGB companion). We suggest it is a member of a class of hybrid Am barium stars which may account for an important fraction of the long missing ``A-type barium stars''. 

The formation of $\gamma$ Cap probably involved a phase of Roche lobe overflow by the AGB companion that stripped part of its envelope and allowed the formation of a WD somewhat lighter than its progenitor mass would have created in isolation. More informative constraints on the progenitor system can be obtained by improving the precision in the WD mass. This can be achieved by obtaining more radial velocity measurements and potentially by an astrometric orbit with smaller errors in the imminent Gaia DR4 release. 

\section*{Acknowledgments}

This research has made use of the CDS Astronomical Databases SIMBAD and VIZIER, NASA's Astrophysics Data System Bibliographic Services, NumPy \citep{van2011numpy} and matplotlib, a Python library for publication quality graphics \citep{Hunter2007}. 

\section*{Data availability}
The data underlying this article are publicly available from the respective archives.

\bibliographystyle{aasjournal}
\bibliography{main}{}

\appendix

\section{Supplementary tables and figures}
\label{app}

Table \ref{table:rv} lists the past radial velocity measurements of $\gamma$ Cap \citep{Campbell1911,Jones1928,Campbell1928} for convenience. 

\begin{figure*}[]
\centering
\includegraphics[width=\textwidth]{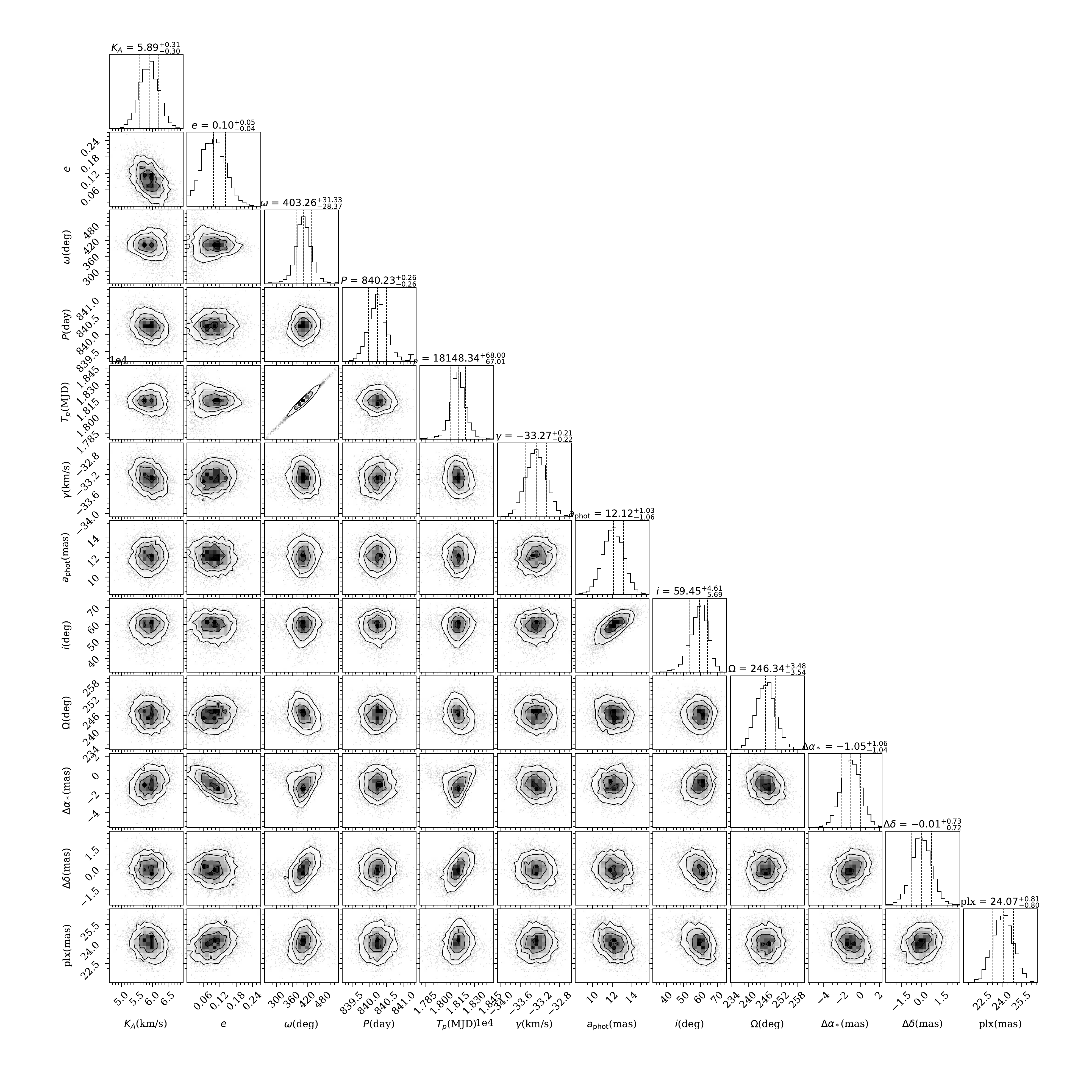}
\caption{\label{fig:orbital_fit_corner} Full orbital and astrometric parameters distributions for Gamma Capricorni.}
\end{figure*}

Figure \ref{fig:orbital_fit_corner} shows the full distributions for the orbital and astrometric parameters of $\gamma$ Cap based on the best fit parameters over 5000 resamples of the data. 

\begin{figure*}[]
\centering
\includegraphics[width=\textwidth]{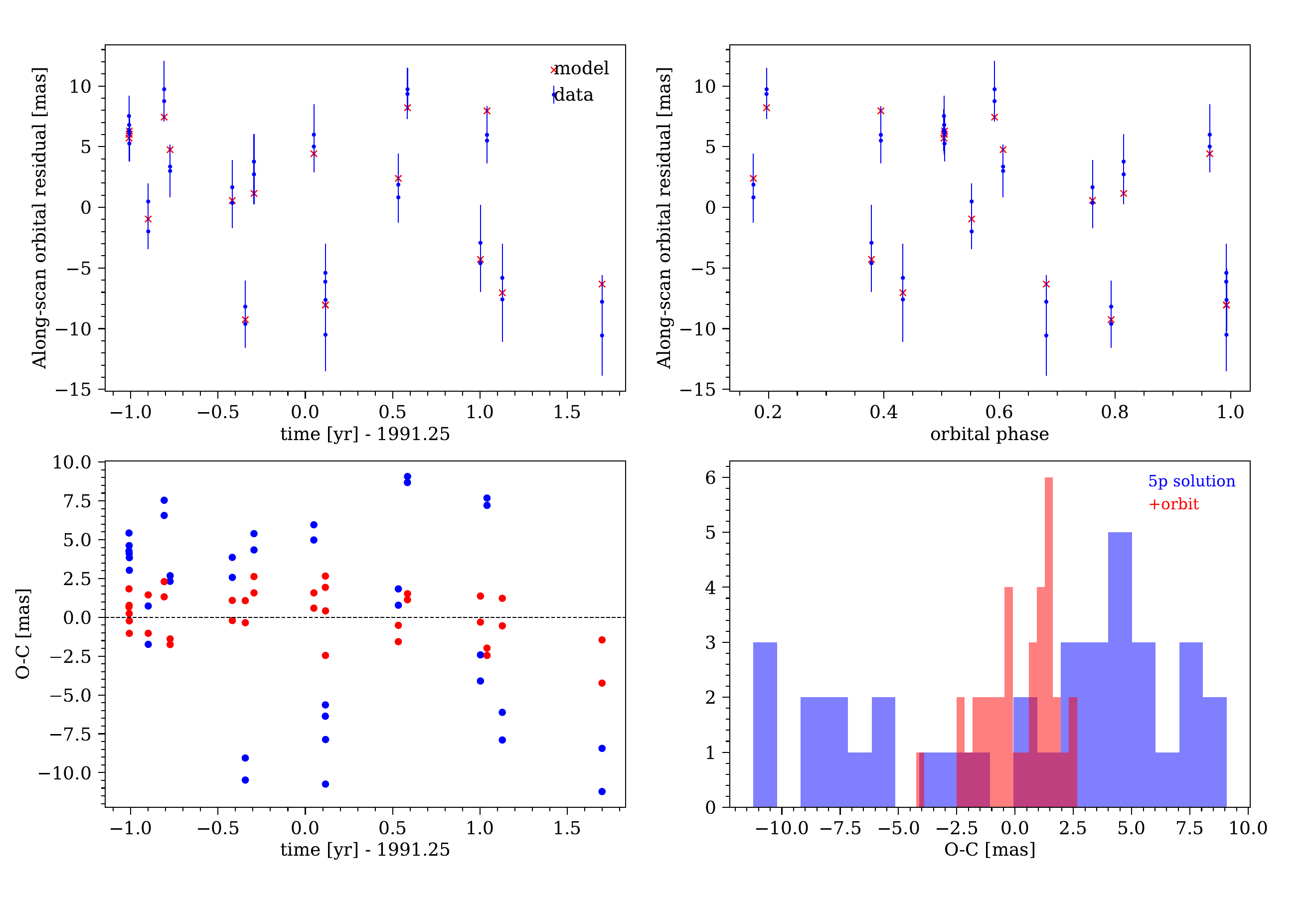}
\caption{\label{fig:hip_details} \textbf{Top left:} Residuals of the Hipparcos 1997 IAD abscissae after subtracting the 5-parameter part of the best fit solution. The red crosses show the corresponding values for the photocenter orbital motion of the best fit solution. \textbf{Top right:} Same as the previous panel but as a function of orbital phase. \textbf{Bottom left:} Corresponding residual with (red) and without (blue) including the photocenter orbit. \textbf{Bottom right:} Same as previous panel but in histogram form.}
\end{figure*}

Figure \ref{fig:hip_details} shows details about the astrometric residuals of the Hipparcos original 1997 IAD. 

\begin{table}
\centering
\caption{\label{table:rv} Past radial velocity measurements of Gamma Capricorni.}
\begin{tabular}{cc}
\hline \hline
JD & $v_{\mathrm{z,A}}$ \\ [0.3cm]

2414854.920 & $-33.5$ \\ [0.3cm]
2414879.866 & $-31.7$ \\ [0.3cm]
2415322.664 & $-35.3$ \\ [0.3cm]
2417400.947 & $-34.6$ \\ [0.3cm]
2417435.862 & $-33.4$ \\ [0.3cm]
2418825.989 & $-26.7$ \\ [0.3cm]
2418889.871 & $-29.7$ \\ [0.3cm]
2422197.83 & $-23.1$ \\ [0.3cm]
2418237.329 & $-34.2$ \\ [0.3cm]
2418250.308 & $-31.1$ \\ [0.3cm]
2419676.367 & $-29.0$ \\ [0.3cm]
2423321.339 & $-36.5$ \\ [0.3cm]
2423324.330 & $-35.5$ \\ [0.3cm]
2423327.316 & $-33.0$ \\ [0.3cm]
2423335.253 & $-33.7$ \\ [0.3cm]
2424027.419 & $-27.4$ \\ [0.3cm]
2424053.336 & $-29.0$ \\ [0.3cm]
2424785.404 & $-28.6$ \\ [0.3cm]

\hline
\end{tabular}
\tablenotetext{0}{Notes:}
\tablenotetext{0}{Sources: \cite{Campbell1911,Jones1928,Campbell1928}.}
\end{table}

\end{document}